# Improved motion correction for functional MRI using an omnibus regression model

Vyom Raval[1,2], Kevin P. Nguyen[1], Cooper Mellema[1], Albert Montillo[1,2]

[1]The University of Texas Southwestern Medical Center
[2]The University of Texas at Dallas

## ABSTRACT

Head motion during functional Magnetic Resonance Imaging acquisition can significantly contaminate the neural signal and introduce spurious, distance-dependent changes in signal correlations. This can heavily confound studies of development, aging, and disease. Previous approaches to suppress head motion artifacts have involved sequential regression of nuisance covariates, but this has been shown to reintroduce artifacts. We propose a new motion correction pipeline using an omnibus regression model that avoids this problem by simultaneously regressing out multiple artifacts using the best performing algorithms to estimate each artifact. We quantitatively evaluate its motion artifact suppression performance against sequential regression pipelines using a large heterogeneous dataset (n=151) which includes high-motion subjects and multiple disease phenotypes. The proposed concatenated regression pipeline significantly reduces the association between head motion and functional connectivity while significantly outperforming the traditional sequential regression pipelines in eliminating distance-dependent head motion artifacts.

## 1. INTRODUCTION

A significant obstacle in the analysis of functional Magnetic Resonance Imaging (fMRI) data is contamination of the blood-oxygenation-level-dependent (BOLD) signal by head motion and fluctuations in non-neuronal physiological processes. Head motion is a particularly significant problem, with even very small movements of the head causing molecular spin history effects that can account for over 90% of fMRI signal [1]. These effects have been shown to produce spurious but structured noise in fMRI scans, causing distance-dependent changes in signal correlation [2]. Since children, the elderly, and diseased patients tend to move more in the scanner than healthy subjects, motion-corruption can act as a major confounding factor leading to erroneous identification of discriminative biomarkers. Consequently, researchers may detect 'biomarkers' of disease which are actually measures of motion artifact instead of true functional connectivity. At the same time, fMRI provides one of the highest, non-invasive spatial resolutions for studying human brain function, making efforts to improve motion artifact suppression highly warranted.

Previous approaches to address this problem have involved pipelines with a sequence of preprocessing steps prior to analysis. Typically, each step consists of a separate motion correction algorithm that often uses linear regression of the data on nuisance covariates, with residuals used in subsequent processing. Lindquist et al. showed that such a sequence of linear filtering operations can reintroduce artifacts removed in prior preprocessing steps [3]. This occurs because each regression step can be modeled as a geometric projection of the data onto a subspace, and a sequence of such projections can project the data into subspaces no longer orthogonal to those previously removed, reintroducing signal related to nuisance covariates. The authors recommended a concatenated regression approach to overcome this problem, with all nuisance covariates combined into a single linear filter applied simultaneously instead of sequentially. However, this finding was largely theoretically motivated, with limited empirical evidence from 21 healthy adults, without the use of well-validated Quality Control (QC) metrics, and without the use of state-of-the-art motion artifact suppression algorithms.

The goal of this study was to develop an omnibus motion suppression pipeline employing concatenated regression that combines state-of-the-art artifact suppression algorithms that have been comprehensively evaluated previously [4] and to quantitatively compare the pipeline to standard sequential regression pipelines. We evaluate the pipelines on a large (n=151) heterogeneous dataset including high-motion, diseased subjects where the results will be most relevant for addressing the motion confounds. We show quantitatively that our new concatenated regression approach suppresses motion artifact, especially the highly problematic, distant-dependent artifact, in resting-state fMRI data better than other widely adopted and popular preprocessing pipelines.

## 2. MATERIALS & METHODS

### 2.1 Dataset

Data used in the preparation of this article were obtained from the Parkinson's Progression Markers Initiative (PPMI) database (www.ppmi-info.org/data). From PPMI, 151 subjects with both functional and anatomical MRI were selected. These subjects included: 1) those with Parkinson's Disease (PD), 2) prodromal subjects, 3) subject scans without evidence of dopaminergic deficiency (SWEDD), 4) and controls, containing a spectrum of motion artifact phenotypes. Summary demographics are shown in **Table 1**.

**Table 1** Demographics for the dataset's 151 subjects.

| Demographic | Value |
| --- | --- |
| Age (years) | 62.8 ± 10 |
| Sex: Male | 71% |
| Cohort | |
| • PD | 67% |
| • Prodromal | 13% |
| • Control | 13% |
| • SWEDD | 8% |
| H&Y Stage of PD patients | 1.7 ± 0.5 |

Imaging was acquired on 3T Siemens scanners. Eyes-open resting-state functional images were acquired using the GE-EPI pulse sequence with TE=25 ms, TR=2400 ms, resolution 68 x 66 x 40 voxels, voxel size 3.294 x 3.294 x 3.3 mm, and scan duration 504 s. T1-weighted anatomical images were acquired using the MPRAGE-GRAPPA sequence with TE=3 ms, TI=900 ms, resolution 176 x 240 x 256 voxels, and 1 mm$^3$ isotropic voxel size.

### 2.2. Preprocessing

Preprocessing followed standard steps for fMRI analysis. Functional images are first processed with FMRIB's Linear Image Registration Tool (MCFLIRT) to perform affine realignment of frames and to compute 6 affine head motion parameters (HMPs) [5]. The realigned images are skull-stripped with FSL Brain Extraction Tool (BET) and Analysis of Functional NeuroImages (AFNI) 3dAutomask [6]. Spatial normali-zation is performed by direct coregistration with an echo planar imaging (EPI) template in Montreal Neurological Institute MNI152 space using the Symmetric Normalization algorithm in Advanced Normalization Tools (ANTs) [7]. Finally, the images are smoothed with a 6 mm full width at the half maximum (FWHM) Gaussian kernel. Anatomical images are skull-stripped with Robust Brain Extraction (ROBEX), spatially normalized to a T1 template with ANTs, and segmented into gray matter/white matter/cerebrospinal fluid with FSL FMRIB's Automated Segmentation Tool (FAST) [5, 7, 8].

### 2.3. Motion correction pipelines

Motion correction is applied to the functional images by regressing out nuisance timeseries from the voxel signals. These nuisance regressors are estimated as follows:

- Head motion parameters (HMPs): 6 rigid-body parameters are computed during frame realignment. The squares and temporal derivatives of these parameters are included for a total of 24 HMPs, as in Friston et al., [1].
- ICA motion components: ICA-AROMA is applied in "nonaggressive" mode to decompose the image into independent components and automatically identify those pertaining to head motion. The number of motion components varies per subject depending on the motion artifact present in the image [9]. This algorithm has been shown to perform favorably against other motion correction strategies [4].
- Physiological (Physio) regressors: the white matter mask is obtained from the anatomical image segmentation, and a mask of non-brain tissue is constructed by computing the complement of the union of the GM and WM segmentations, as described in [4]. These two masks are then eroded as recommended by Power et al. [2] with the method described by Parkes et al [4]. The mean signal from these two masks is used to define the physiological regressors.

Three motion correction pipelines are constructed, each consisting of a distinct ordering of these regression steps.

Let **y** be an $n$-dimensional vector containing the fMRI signal from a specific voxel. Further, let $\mathbf{X}_{HMP}$ be an $n \times 24$ design matrix containing 24 HMP regressors, $\mathbf{X}_{AROMA}$ be an $n \times p$ design matrix containing $p$ ICA-AROMA noise components, $\mathbf{X}_{Physio}$ be an $n \times 2$ design matrix containing 2 Physio regressors.

The effects of these nuisance regressors are removed from **y** by fitting a linear model of the form:
$$\mathbf{y} = \mathbf{X}\boldsymbol{\beta} + \mathbf{e}$$
And the residual **e** is used for further analysis.

Two pipelines conducted the regression steps sequentially:

1) HMP > AROMA > Physio
$$\mathbf{e} = ((\mathbf{y} - \mathbf{X}_{HMP}\boldsymbol{\beta}_1) - \mathbf{X}_{AROMA}\boldsymbol{\beta}_2) - \mathbf{X}_{Physio}\boldsymbol{\beta}_3$$

2) AROMA > HMP > Physio
$$\mathbf{e} = ((\mathbf{y} - \mathbf{X}_{AROMA}\boldsymbol{\beta}_4) - \mathbf{X}_{HMP}\boldsymbol{\beta}_5) - \mathbf{X}_{Physio}\boldsymbol{\beta}_6$$

where ">" indicates the sequence of regression steps.

The third pipeline implements our proposed simultaneous regression approach, in which the three sets of nuisance regressors are concatenated and regressed out of the image simultaneously. This pipeline is referred to as:

3) [AROMA, HMP, Physio]
$$\mathbf{e} = \mathbf{y} - [\mathbf{X}_{HMP}\mathbf{X}_{AROMA}\mathbf{X}_{Physio}]\boldsymbol{\beta}_7$$

where the [ ] symbol indicates concatenation of nuisance regressors into a single matrix.

### 2.4. Quality assessment

A standard resting-state functional connectivity analysis is conducted on the motion corrected images. The Gordon functional atlas is applied to parcellate the brain into 333 cortical regions-of-interest [10]. The functional connectivity (FC) matrix is constructed for each subject by computing the

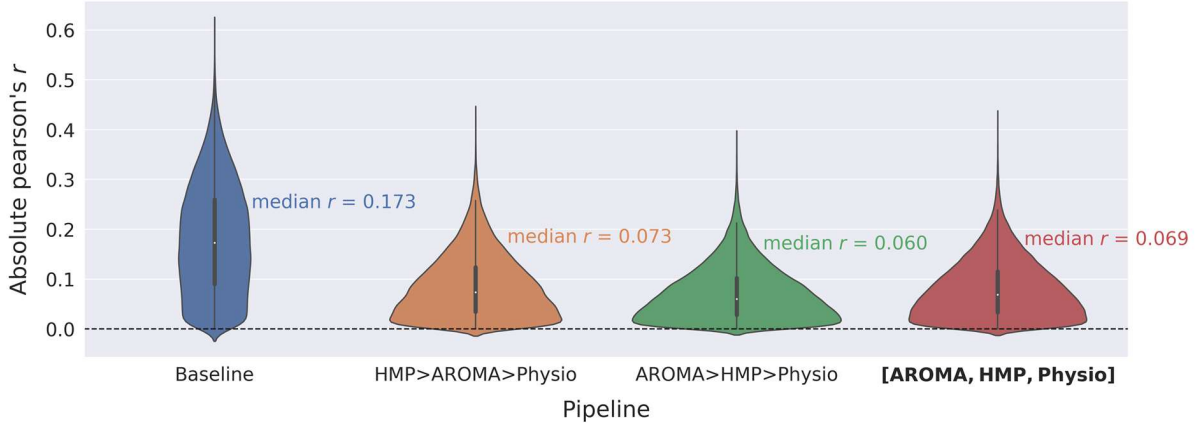

**Figure 1** Distribution of <u>QC-FC correlations</u> of the 55,278 FC matrix edges for each pipeline. The median absolute Pearson's *r* is annotated. Greater values indicate greater contamination of FC matrices with motion artifact. Our proposed pipeline is shown in red and bolded.

Pearson's correlation coefficient between the mean regional fMRI timeseries of each pair of regions.

Next, each subject's head motion is quantified by computing framewise displacement from their HMPs. The framewise displacement at a timepoint *t* is defined by Power et al. as:

$$FD(t) = |d_x(t) - d_x(t-1)| + |d_y(t) - d_y(t-1)|$$
$$+ |d_z(t) - d_z(t-1)| + |\theta_x(t) - \theta_x(t-1)|$$
$$+ |\theta_y(t) - \theta_y(t-1)| + |\theta_z(t) - \theta_z(t-1)|$$

where $d_x, d_y, d_z$ are the rigid-body translation parameters and $\theta_x, \theta_y, \theta_z$ are the rotation parameters [11]. Rotation parameters are converted from radians to millimeters by multiplying by an approximate brain radius of 50 mm, following the approach of Power et al. [11].

Following the work of Parkes et al., two quality metrics are computed from the FC matrix and mean framewise displacement (mFD) of each subject [4]. For each of the 55,278 edges in the FC matrix, the <u>QC-FC correlation</u> is computed, which is the Pearson's correlation between mFD and the FC value of that edge among the subjects. QC-FC correlation values close to 0 are desirable, indicating that FC edge values are not correlated with motion. Distance-dependent motion artifact over the entire FC matrix is quantified using <u>QC-FC distance dependence</u>, which is the Spearman's rank correlation coefficient between the QC-FC correlation of each edge and the Euclidean length of the edge in the brain. Spearman's correlation is used to capture the possible non-linearity of associations between QC-FC and inter-regional distance. Head motion has been shown to inflate the values of shorter functional connections relative to longer connections, and consequently distance dependence correlation closer to 0 is desired [2, 4, 11]. As an improvement to previous QC evaluations and to facilitate comparisons, the QC metrics are also computed for a 'Baseline' FC obtained from our images without applying any motion correction.

## 3. RESULTS

We first examine the efficacy of different noise correction pipelines in removing the confounding relationship between in-scanner movement and estimates of FC, as assessed with <u>QC-FC correlation</u>. The distributions of these correlations revealed that no pipeline was able to *completely* reduce the median absolute QC-FC correlation to zero. However, each motion correction pipeline performed comparably and significantly reduced the median absolute QC-FC correlation compared to baseline (**Fig. 1**).

Next, we examine the residual association between QC-FC and inter-regional distance with the absolute <u>QC-FC distance dependence</u>, where lower values indicating less remaining dependence are better. Without motion suppression there is distance dependence of $rho = 0.10$ (baseline, blue), while the proposed omnibus pipeline (red) achieves far lower distance dependence ($rho = -1.15 \times 10^{-5}$) compared to the sequential pipelines (**Fig. 2**, orange, green). The statistical significance of the association between inter-regional distance and QC-FC was computed for each pipeline, with more significant associations (lower p-values) indicating *greater* distance dependent artifacts in the FC matrices. ***Only the proposed pipeline successfully eliminated significant distant-dependent artifact (p = 0.998) substantially outperforming the sequential pipelines*** both of which retained significant distance-dependence (p = 5.07×10⁻⁵ and p = $1.89 \times 10^{-21}$).

## 4. DISCUSSION

Head motion artifact is a pernicious problem in fMRI data analysis, often contaminating the neural signal of interest leading to spurious group differences while masking true differences. Our results demonstrate that without any motion correction preprocessing, there is high corruption of FC (absolute median QC-FC correlation $r = 0.17$, absolute distance-dependent correlation $rho = 0.10$). We also show that while both concatenated and sequential regression pipelines are able to substantially reduce QC-FC correlations, only concatenated regression is successful in removing distance-dependent artifacts in the FC (*rho* = -1.15×10$^{-5}$ and p = 0.998 significance demonstrating that inter-regional distance and QC-FC are uncorrelated). Notably, while the AROMA>HMP>Physio pipeline had slightly better QC-FC correlations (**Fig. 1**) compared to the proposed pipeline, it showed significantly negative (non-zero) distance-dependent QC-FC associations (**Fig. 2**), indicating that the dependence of FC values on motion becomes more negative with longer inter-regional distances, a pattern seen before [2, 4, 11]. Only the proposed pipeline is able to eliminate these associations.

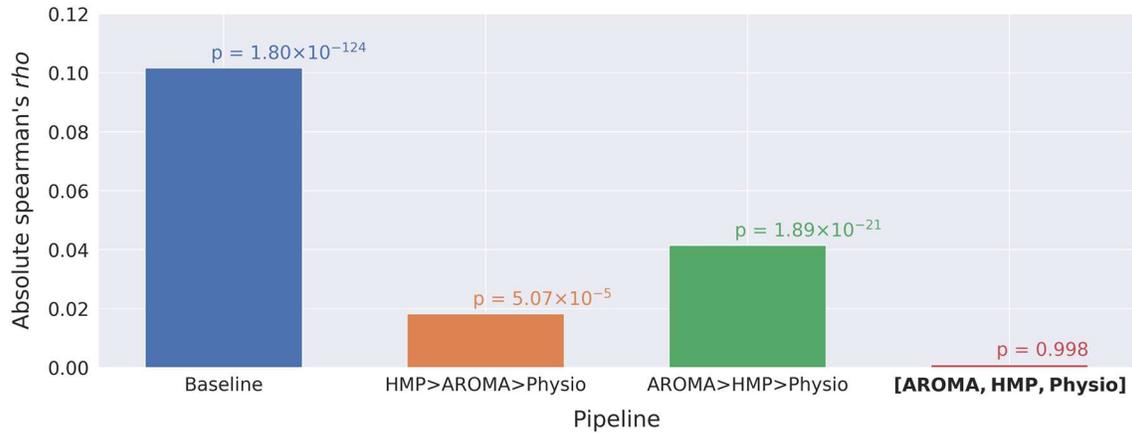

**Figure 2** Values of <u>QC-FC distance dependence</u> for each pipeline. The p-value indicates the significance of the association between inter-regional distance and QC-FC, with more significant associations (lower p-values) indicating greater distance dependent artifacts in the FC matrices. Our proposed pipeline is shown in red and bolded.

It remains unclear whether, even after extensive preprocessing, motion artifact is ***completely*** removed from the FC. This is because there is no 'ground-truth' for FC against which we can validate motion corrected images. Future work could potentially address this by simulation experiments where spin-history motion artifacts are introduced to low motion fMRI data. We also plan to develop additional quality control metrics based on multivariate correlations between FC matrix values and motion, which may further reveal the extent of motion artifact present in the FC. Finally, we plan to evaluate the application of our pipeline on task-based fMRI, where improved motion artifact correction may reduce spurious brain activation signals and enhance the detection of true neuronal activity.

## 5. CONCLUSION

This work proposes a new omnibus motion correction pipeline for fMRI that significantly reduces more motion artifact compared to standard sequential pipelines. The proposed pipeline can be directly applied to help reduce confounds in rs-fMRI analyses including FC analyses, and thereby improve the detection of true discriminative biomarkers. As demonstrated on a variety of movement disorder phenotypes, the proposed approach is disease agnostic and should be applicable to a wide variety conditions plagued by motion including other neurodegenerative, neurodevelopmental, and mental disorders.

## ACKNOWLEDGEMENTS

PPMI – a public-private partnership – is funded by the Michael J. Fox Foundation for Parkinson's Research and funding partners, including Abbvie, Allergan, Avid Radiopharmaceuticals, Biogen, Biolegend, Bristol-Myers Squibb, Celgene, Denali, GE Healthcare, Genentech, GlaxoSmithKline, Lilly, Lundbeck, Merck, Meso Scale Discovery, Pfizer, Piramal, Prevail Therapeutics, Roche, Sanofi Genzyme, Servier, Takeda, Teva, UCB, Verily, and Voyager Therapeutics.